\documentclass[preprint,journal]{vgtc}            

\onlineid{0}

\preprinttext{
    \parbox{\textwidth}{\centering%
        $\copyright$ \the\year~IEEE. This is the author's version of the article that has been accepted by IEEE VIS 2026\\and will be published in IEEE Transactions on Visualization and Computer Graphics.%
        }%
    }

\vgtccategory{Research}

\title{ManyFold: A Design Exploration of Data Visualization on Foldable Mobile Devices}

\author{%
  \authororcid{Julian Baader}{0009-0009-7585-459X},
  \authororcid{Ricardo Langner}{0000-0003-4519-2168},
  \authororcid{Can Liu}{0000-0003-3267-3317},
  \authororcid{Raimund Dachselt}{0000-0002-2176-876X}
  }

\authorfooter{
  \item
  	J. Baader, R. Langner, and R. Dachselt are with the Interactive Media Lab Dresden, TU Dresden, Germany.
  	E-mail: {firstname}.{lastname}@tu-dresden.de. 
    
   \item
   	R. Dachselt is also with the Centre for Tactile Internet with Human-in-the-Loop (CeTI).

  \item C. Liu is with the School of Creative Media, City University of Hong Kong, China.
  	E-mail: {firstnamelastname}@cityu.edu.hk.

  \item 
  C. Liu and R. Dachselt are corresponding authors.  
}

\abstract{With this work, we explore the unique potential of data visualization on novel foldable mobile devices (foldables).
Even though foldables are already commercially available, there is limited knowledge of how to leverage their distinct characteristics for visualization.
This gap will only grow as their form factors become increasingly diverse.
To address this, we use a two-step approach.
First, we present a device-centered design space, structured around physical and usage properties of foldable devices.
Second, we introduce a conceptual framework that investigates  visualization on foldables from four complementary perspectives: 
    \textit{More Displays} -- distributing multiple views to leverage additional display space;
    \textit{More Shapes} -- mapping visualizations to spatial fold states;
    \textit{More Interactions} -- coupling visualization tasks and folding interactions; and
    \textit{More States} -- enabling responsive visualization through folding.
We complement the design space and conceptual framework with a low-fidelity ideation workshop and the subsequent prototyping of interactive artifacts.
We reflect on the results and lessons learned from our exploratory, design-driven process and discuss opportunities and challenges, including the gap between our proposed concepts and commercially available foldable devices.
By providing conceptual foundations and illustrating the potential of foldables, we hope to inspire and inform the development of future visualization applications for this evolving class of devices.}


\teaser{
    \centering
    \includegraphics[width=\linewidth]{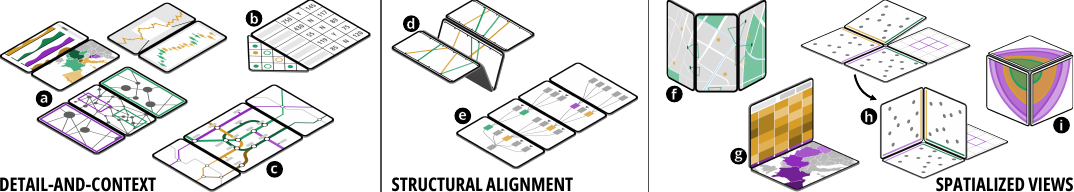}
    \caption{
        Part of our conceptual framework: \emph{detail-and-context techniques} (left), the \emph{structural alignment} of visualizations (middle), and \emph{spatialized views} (right);
        \figureLabel{a} Different overview+detail settings;
        \figureLabel{b} Adjacency matrix on a triangular panel;
        \figureLabel{c} Focus and context on separate panels;
        \figureLabel{d} Parallel axes mapped to panel edges;
        \figureLabel{e} Separate levels of a DOI graph;
        \figureLabel{f} Perspective wall on tri-fold;
        \figureLabel{g} Temporal and spatial data on orthogonal panels;
        \figureLabel{h} Scatterplot matrix on partial cube;
        \figureLabel{i} Hive plot in a volumetric configuration.
    }
    \label{fig:detail-context-shapes}
}

\graphicspath{{figs/}{figures/}{pictures/}{images/}{./}} 

\usepackage{tabu}                      
\usepackage{booktabs}                  
\usepackage{lipsum}                    
\usepackage{mwe}                       
\usepackage{microtype}
\usepackage{svg}

\usepackage{siunitx}
\usepackage[dvipsnames]{xcolor}
\usepackage{ccicons}
\usepackage{xspace}

\newcommand{\etal}[1]{{#1}~et~al.\xspace} 
\newcommand{\eg}{e.g.,\xspace}

\definecolor{colorCompositionText}{HTML}{5F6ED3}
\definecolor{colorUsageText}{HTML}{50B151}
\definecolor{colorCompositionHighlight}{HTML}{C0C5EA}
\definecolor{colorUsageHighlight}{HTML}{C1DEBD}
\newcommand{\compositionDim}[1]{\colorbox{colorCompositionHighlight!50}{\textit{#1}}}
\newcommand{\usageDim}[1]{\colorbox{colorUsageHighlight!50}{\textit{#1}}}

\newcommand{\mymarginnote}[1]{\ifinner\else\marginpar{\footnotesize\raggedright #1}\fi}

\newcommand{\figureRef}[2][]{\hyperref[#2]{\autoref*{#2}#1}}

\newcommand{\examples}{\par\textbf{Examples:}\xspace}
\newcommand{\figureLabel}[1]{(#1)\xspace}

\newcommand{\smallParagraph}[1]{\par\textbf{#1:}\xspace}

\usepackage{mdframed}

\newmdenv[
    topline=false,
    bottomline=false,
    rightline=false,
    linewidth=1pt,
    linecolor=Mulberry,
    innerleftmargin=4pt,
    innerrightmargin=2pt,
    innertopmargin=2pt,
    innerbottommargin=2pt,
    leftmargin=0pt,
    rightmargin=0pt,
    skipabove=2pt,
]{textbox}

\newcommand{\sectionEnd}[1]{\begin{textbox}#1\end{textbox}}
\newcommand{\idea}[1]{{\sffamily\small\bfseries #1}}

\usepackage{mathptmx}                  

\begin{document}



\maketitle
\section{Introduction}
Mobile devices have become a predominant computing platform.
Beyond traditional flat-screen designs, recent technological advances have introduced novel mobile form factors.
Most notably, several manufacturers are working to bring foldable mobile devices to the mass market, with commercial success to date. 
Further developments include rollable or flexible screens (\eg \cite{Steimle:Flexpad:2013,Butscher:InformationSense:2017,Khalilbeigi:Xpaaand:2011,Gomes:MagicScroll:2018}) that promise to expand the possibilities of mobile computing.
These developments can be understood within the broader category of shape-changing interfaces, where digital devices alter their physical configuration~\cite{rasmussen2012,sturdee2019}.
Unlike rollable or continuously flexible displays, foldables provide a limited set of discrete and stable form factors.
The compact form factors promise enhanced mobility while offering larger screen spaces when necessary.
Two-panel configurations dominate current offerings, while three-panel designs are emerging.
Ongoing advancements in display and hinge technologies further suggest a future of increasingly diverse and complex variants.
Meanwhile, research on foldables has demonstrated their potential for collaborative work~\cite{Saniee-monfared:TentMode:2020,Hinckley:Codex:2009}, 3D content manipulation~\cite{Bueschel:Foldable3D:2016,Teetz:FoldAR:2024,Can:AngleCAD:2022}, and adaptive user interfaces that respond to physical device configurations~\cite{khaddam2020,Gomes:PaperFold:2015}.
These approaches also demonstrate folding as a new input and interaction modality, which extends to wearable devices, such as wrist-worn displays and smartwatches \cite{Klamka:Watch+Strap:2020,Fuchs:FoldWatch:2018,Zhu:WristOrigami:2018}.

In our increasingly data-driven world, visualization has become essential beyond desktop contexts~\cite{roberts2014}, including mobile devices, particularly for visualizing personal information.
Mobile data visualization has grown as a research area, with dedicated research events~\cite{watson2015,lee2018,DagstuhlSeminar2019} and a recent book providing an overview of the field~\cite{MobileDataVisualization2021}.
However, mobile visualization remains constrained by the limited screen space and the associated interaction issues~\cite{chittaro2006,langner:book-intro:2021}.
Despite high resolutions and pixel densities, visualization elements cannot be downsized arbitrarily without compromising visibility and readability~\cite{richer-2024-ScalabilityVisualization,robertson-2009-ScaleComplexityVisual}.
As a consequence, established techniques such as multiple coordinated views~\cite{roberts1998encouraging} are difficult to transfer to mobile contexts.
While responsive design strategies~\cite{Horak:Book-Responsive:2021,hoffswell2020} aim to address such issues, they still assume a single, flat, and rectangular display with a fixed size.

We, therefore, argue that foldables represent a promising opportunity for visualization research.
However, it remains unclear how foldable devices can best be utilized specifically for visualization purposes.
In particular, we do not yet know:
    Which visualizations are especially well-suited for foldable form factors?
    Can the unique properties of foldable devices inspire entirely new visualization approaches?
    How might folding actions enhance data analysis tasks?

Our work aims to fill the gap exemplified by these questions, which requires an understanding of both the device characteristics and their visualization potential.
To this end, we first present a device-centered design space that systematically describes their physical composition and usage dimensions, looking beyond existing products.
Building on this foundation, we detail a conceptual framework focused on the use of visualizations that leverage the unique physical properties and interaction capabilities of foldables.
We highlight the manifold possibilities by covering common visualization tasks and techniques and explore concepts from four complementary perspectives:
    \textit{More Displays} -- distributing multiple views to take advantage of additional display space;
    \textit{More Shapes} -- utilizing spatial fold states for visualization;
    \textit{More Interactions} -- coupling visualization tasks and folding; and
    \textit{More States} -- employing responsive visualization through folding.
We complement our design space and the conceptual framework with prototyping activities.
Specifically, we conducted an ideation workshop with 15 students to envision and explore concrete visualization concepts and use cases.
We further describe four approaches to develop and explore interactive visual artifacts, making our concepts more tangible and taking a step towards concrete prototypes.
Reflecting on both our conceptual and prototyping efforts, we discuss limitations, implications, and opportunities for visualizing data on foldable devices.

In summary, our main contributions are:
    (i)\,A design space of the device composition and usage characteristics of foldables, applicable beyond data visualization.
    (ii)\,A conceptual framework that investigates visualization on foldables from four complementary perspectives, yielding a set of exemplary visualization concepts.
    This is completed by:
    (iii)\,Findings from a paper prototyping workshop and initial interactive prototypes.
    (iv)\,A discussion of the opportunities and challenges of visualizing data on foldables.
\section{Related Work}
\label{sec:rw}
In the following, we review prior work on mobile device form factors, mobile data visualization, and the use of folding within visualization.
We further ground this work in current commercial foldable devices.

\paragraph{Novel Mobile Devices and Interactions}%
\label{sec:rw:novel}
Form factors beyond conventional flat screens have attracted attention in both research and industry for more than two decades~\cite{mone2013}.
In particular, bendable, rollable, foldable, and composed device configurations have been explored, along with the post-WIMP interactions they enable.
Much of this development has been driven by advances in deformable display technologies (\eg paper-like or stretchable), with \etal{Kim}~\cite{kim2023} positioning foldables at the beginning of a roadmap toward fully deformable displays.

Research on novel mobile form factors has progressed from limited bending to fully reconfigurable devices.
Early work, such as Gummi~\cite{schwesig2004}, proposed bending as a direct interaction technique, and subsequent research extended this with active feedback~\cite{Strohmeier:ReFlex:2016} and further highlighted the value of bending for spatial interaction.
Other works investigate even more flexible, paper- or cloth-like devices that support bend gestures (such as bending a corner of the device)~\cite{Lahey:PaperPhone:2011,Girouard:One-Hand-Bend:2015} or even complete folding~\cite{Steimle:Flexpad:2013,Butscher:InformationSense:2017}.
For instance, InformationSense~\cite{Butscher:InformationSense:2017} uses a deformable cloth display to allow folding of an information space.
A related direction explores rollable devices reminiscent of paper scrolls, which change their available display area by extending or retracting a flexible screen~\cite{Khalilbeigi:Xpaaand:2011,Steimle:CHI-EA:2012,Gomes:MagicScroll:2018}.
In parallel, folding along fixed axes or hinges has been explored as a way to combine multiple displays into reconfigurable mobile devices~\cite{Hinckley:Codex:2009,Khalilbeigi:FoldMe:2012,Can:AngleCAD:2022}. 
Paddle~\cite{ramakers2014} uses a more dynamic folding approach with a reconfigurable device inspired by the Rubik's Magic puzzle.
Moving from research prototypes to commercial products, current foldables fall into a few dominant form-factor categories:
    book-style foldables unfold from a phone into a small tablet, 
    clamshell devices fold a phone-sized display in half, and, more recently, 
    tri-fold devices offer even greater display extension.
Most commercial foldables include a cover display usable while folded, which has grown progressively larger in recent models.
An overview of representative current devices is provided in the supplemental material.

Foldable devices introduce new interaction opportunities by changing how display surfaces are arranged and manipulated.
Folding itself has been explored as an input modality, through continuous control (\eg zooming~\cite{Bueschel:Foldable3D:2016}) or discrete fold gestures~\cite{yeh2024}.
The multi-panel form factor also affects touch interaction: prior work has examined how fold angle and holding posture affect touch performance on angled surfaces~\cite{hennecke2012,yamakawa2024}, proposed gestures that cross hinges or move content between panels~\cite{Li:TouchGestures:2022,Shen:Pull-Gestures:2022}, and explored the hinge itself as an input area~\cite{Perelman:HingeGestures:2026}.
The ability of foldables to transform between fold states has also been explored.
Specific states can trigger viewport changes or mode switches, as demonstrated by PaperFold~\cite{Gomes:PaperFold:2015}, where folding into a triangular hull configuration activates a 3D view, or by \etal{Spindler}~\cite{Spindler:Palettes:2010}, where unfolding a tangible palette reveals additional user interface elements.
More generally, spatial fold states have been used to support tasks that benefit from a non-planar arrangement of screens, including 3D navigation and modeling~\cite{Bueschel:Foldable3D:2016,Can:AngleCAD:2022}, as well as AR object manipulation~\cite{Teetz:FoldAR:2024}.
These spatial arrangements can also support collaboration by creating shared-view or face-to-face setups~\cite{Hinckley:Codex:2009,Saniee-monfared:TentMode:2020}, illustrating how foldable devices extend beyond handheld use to surface-supported and even wearable form factors~\cite{Fuchs:FoldWatch:2018,Zhu:WristOrigami:2018}.

Various prototyping technologies have been used to realize these novel form factors.
Many early works used deformable surfaces, projectors, and depth-sensing cameras~\cite{Steimle:Flexpad:2013,Butscher:InformationSense:2017,ramakers2014}.
Others employed e-paper displays~\cite{Gomes:PaperFold:2015} or commercial mobile devices~\cite{Bueschel:Foldable3D:2016,Can:AngleCAD:2022} to build modular, physically reconfigurable devices.
Today, flexible OLED technology has made foldable form factors commercially viable.
Beyond physical prototyping, Flecto~\cite{khaddam2020} provides a model-based approach for simulating foldable user interfaces in a virtual 3D environment.

\paragraph{Mobile Data Visualization}
\label{sec:rw:mobilevis}
Data visualization research extends beyond traditional desktop settings to a wide range of display environments, from large shared displays~\cite{Belkacem:Large-displays:2024} to small personal devices~\cite{roberts2014,lee2018,chittaro2006,Klamka:Watch+Strap:2020}.
Mobile devices such as smartphones, tablets, and smartwatches have received particular attention~\cite{MobileDataVisualization2021}.
A central challenge in mobile visualization is the limited and fixed display space, which has motivated work on glanceable visualizations for small screens~\cite{Blascheck:Glanceable:2019} and on responsive visualization, where layout, information density, and interaction adapt to specific device characteristics~\cite{hoffswell2020,Horak:Book-Responsive:2021}.
These constraints become especially apparent with multiple coordinated views, which are difficult to accommodate on a single small screen~\cite{sadana2016}.
To overcome the constraints of individual mobile screens, several approaches combine multiple devices into visualization workspaces.
VisTiles~\cite{Langner:VisTiles:2018} demonstrates this by coordinating multiple tablets for visual data exploration, where the spatial arrangement of devices and their side-by-side placement drive how views are distributed and combined.
Building on this, Vistribute~\cite{Horak:Vistribute:2019} formalizes the matching of visualizations to device characteristics.
This direction has also since been extended by combining mobile devices with AR~\cite{Langner:Marvis:2021}.

Both responsive and multi-device approaches address the display limitations of mobile devices. 
Foldable devices relate to responsive visualization, since their visible display area changes with folding, and to cross-device visualization, as their reconfigurable panels echo multi-display environments within a single portable form factor.
While prior work has addressed 3D content interaction~\cite{Bueschel:Foldable3D:2016,Can:AngleCAD:2022} and map interface adaptation~\cite{Savino:Maps:2024} for foldables, to our knowledge, no prior work has explored data visualization on foldables.

\paragraph{Folding in and for Visualization}
\label{sec:rw:foldvis}

\begin{figure*}
    \centering
    \includegraphics[width=\linewidth]{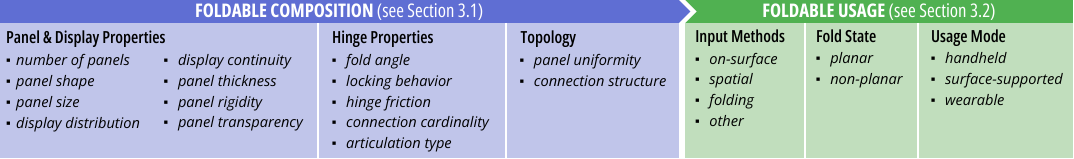}\textbf{}
    \caption{
        Our design space for foldable devices, organized into six categories across two groups. Foldable composition (\textcolor{colorCompositionText}{blue}) defines what a device is.
        Foldable usage (\textcolor{colorUsageText}{green}) defines how it is used. The composition influences and constrains the usage.
    }
    \label{fig:design-space}
\end{figure*}

Independent of hardware, folding and related physical transformations have served as both a metaphor and an interaction principle in visualization.
\etal{Bludau}~\cite{Bludau:FoldableDataVis:2025} surveyed techniques under the umbrella of \emph{un/foldable visualizations} to fluidly reveal or reduce information.
They show that many visualization techniques rely on operations that can be understood as a form of folding: expanding and collapsing hierarchies or adding detail through semantic zooming.
Some techniques simulate a folded or angled three-dimensional surface arrangement as part of their visual encoding.
For example, the perspective wall~\cite{mackinlay1991} arranges linear information along angled surfaces receding from a central focus region.
Similarly, Roller~\cite{Wang:Roller:2007} uses multiple angled faces to present more information than a flat layout.
Other techniques use folding as an explicit interaction mechanism~\cite{elmqvist2010,Butscher:SpaceFold:2014}.
For example, \etal{Tominski}~\cite{Tominski:FoldingComparison:2012} let users fold away overlapping views to reveal and compare underlying data, similar to peeling back a page.
M\'elange~\cite{elmqvist2010} and SpaceFold~\cite{Butscher:SpaceFold:2014} fold away regions to bring areas of interest closer together, and \etal{Chiu}~\cite{Chiu:Document-folding:2011} use multi-touch gestures to fold virtual documents.
A close example of coupling physical and conceptual folding is InformationSense~\cite{Butscher:InformationSense:2017}, where users physically fold a deformable cloth display to fold the underlying map visualization.
We believe that foldable mobile devices can offer a similar opportunity, in which folding or unfolding the device can trigger corresponding visualization operations.
\section{Design Space of Foldables}
\label{sec:foldable-design-space}

\begin{figure}[b]
    \centering
    \includegraphics[width=\linewidth]{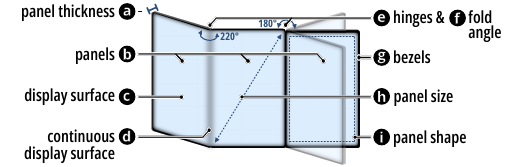}
    \caption{Key physical characteristics of a foldable device (a--i), shown for a three-panel configuration with rectangular panels.}
    \label{fig:physical-dimensions}
\end{figure}

The ability to change the physical configuration of foldable mobile devices affects how visual content can be displayed, perceived, and interacted with.
To better understand these implications, we first need to develop a thorough understanding of the foldable composition (\ref{subsec:composition}) and its usage characteristics (\ref{subsec:usage-properties}).
To this end, we present a device-centered design space, constructed through iterative discussions among the authors.
We began by surveying foldable and related reconfigurable mobile interfaces, spanning commercial products and research prototypes as well as foldable artifacts beyond computing, such as physical folding objects and paper-folding techniques.
We then discussed these examples, identified recurring properties, and organized them into initial dimensions.
Through further discussion, we refined these dimensions and generatively extended them to also account for configurations that are technically conceivable but not yet realized, until we reached consensus.
The resulting design space is organized into six categories, as illustrated in \figureRef{fig:design-space}.
Although this paper's primary focus is on visualization on foldables, the dimensions we describe still apply to a range of interactive applications and thus can inform the general design of user interfaces for foldable hardware.

\subsection{Foldable Composition}%
\begin{figure}[b]
    \centering
    \includegraphics[width=\linewidth]{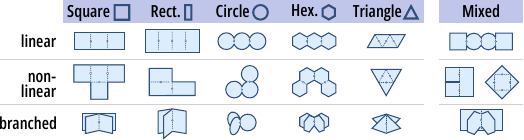}
    \caption{
        Examples of foldable device \compositionDim{connection structures} (rows) with different \compositionDim{panel shapes} and \compositionDim{panel uniformity} (columns).
    }
    \label{fig:composition}
\end{figure}

\begin{figure}[t]
    \includegraphics[width=\linewidth]{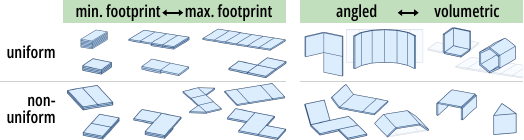}
    \caption{
        Foldables with uniform and non-uniform panels, in planar (min. to max. footprint) and non-planar (angled to volumetric) fold states.
}
    \label{fig:states}
\end{figure}

\label{subsec:composition}
We refer to the physical properties that make up a foldable as the device's composition.
It defines the possible ways the device can fold and unfold, the resulting spatial states it can form, and how it can be used.
The foldable composition is defined by the panel and display properties, the hinge properties, and the device's topology (see \figureRef{fig:physical-dimensions}).
We provide a set of example compositions in \figureRef{fig:composition} and \ref{fig:states}, chosen not to be exhaustive but to reflect plausible configurations of future foldables.

\paragraph{Panel \& Display Properties}
The fundamental building blocks of a foldable device are \emph{panels} (\figureRef[b]{fig:physical-dimensions}).
Each panel can be connected to other panels via mechanical hinges (\figureRef[e]{fig:physical-dimensions}), enabling folding or unfolding.
Importantly, panels are distinct from displays and could also serve as non-visual structural components.
Foldable devices may consist of two or more distinct panels (\compositionDim{number of panels}).
Two-panel devices support only basic folding (\eg book-style or clamshell), whereas three or more panels enable more complex spatial states.
Panels can differ in \compositionDim{shape} and \compositionDim{size} (\figureRef[i+h]{fig:physical-dimensions}).
Future foldables might feature non-rectangular panel shapes such as circles, triangles, or other specialized geometries~\cite{nonaka2008}.
Although such forms introduce constraints on hinge design and complicate content layout, they could benefit specialized applications like wearable gadgets and textile interfaces~\cite{peetz2019-bodyhub}.
Several examples of panel shapes and how they affect the composition of a foldable can be seen in \figureRef{fig:composition}.
The \compositionDim{size} of panels also affects the usage properties of a foldable.  

A panel can, in principle, have one or more individual displays (\compositionDim{display distribution}).
Particularly relevant is whether a panel has a display on one or both sides, as this affects how the device is held and used.
Since a display does not necessarily cover the full panel surface, bezels may appear along panel edges (\figureRef[g]{fig:physical-dimensions}).
This can lead to gaps in the display surface, which can segment content~\cite{wallace2014} and cause visual occlusion or distortions~\cite{dealmeida2012}, similar to tiled displays and multi-monitor setups.
When a single flexible display spans multiple panels (\compositionDim{display continuity}), bezels at hinges may be eliminated, though a noticeable crease or edge may remain (\figureRef[d]{fig:physical-dimensions}).

Finally, advances in display and related technologies are influencing three further properties.
The \compositionDim{panel thickness} (\figureRef[a]{fig:physical-dimensions}) in both the folded and unfolded states affects handling, perception, and supported fold types.
Thinner panels could enable paper-like flexibility, while thicker panels provide rigidity for comfortable handling (\compositionDim{panel rigidity}).
These properties do not need to be uniform across a device, as rigid and flexible panels can be combined (\eg \cite{rendl2016}).
For foldables, the transparency of displays and panels could also offer unique benefits (\compositionDim{panel transparency}).
We see rich opportunities for systems using (semi-)transparent displays to support superimposed digital content, i.e., multiple display layers that visually align.

\paragraph{Hinge Properties}
Hinges connect neighboring panels and define the range and direction of possible folds (\figureRef[e]{fig:physical-dimensions}).
A key characteristic is the \compositionDim{maximum fold angle}, i.e., how far connected panels can fold relative to each other (\figureRef[f]{fig:physical-dimensions}).
It determines which fold states a device can reach: a maximum of \ang{180} permits flat layouts, while a larger range enables folding in both directions around the hinge axis.
Devices with a continuous display spanning multiple panels (\compositionDim{display continuity}) place the hinge behind the display, which restricts folding to a single direction.
Hinges further differ in their behavior at specific angles (\compositionDim{hinge locking behavior}).
They may allow continuous movement, holding any intermediate angle, or provide discrete locking points that stabilize particular fold states.
The \compositionDim{hinge friction} also influences handling; low-friction hinges support easier one-handed operation, whereas higher friction may require two hands.

Furthermore, a single hinge may connect more than two panels (\compositionDim{hinge connection cardinality}).
While constrained by factors like \compositionDim{panel thickness}, these designs can enable complex geometries such as fan-like structures (see branched connection structures in \figureRef{fig:composition}).
Also, hinges may support different \compositionDim{articulation types} beyond simple folding, such as swiveling, pivoting, sliding, or detaching.
This, in turn, constrains the physical form of the hinge (\eg line hinge or point joint hinge).
For example, Gomes and Vertegaal~\cite{Gomes:PaperFold:2015} study detachable panels and their effects on viewport transformations.
However, detaching falls outside the scope of folding as defined here.

\paragraph{Topology}
We refer to the physical arrangement of panels that make up a foldable as the device's topology.
It is defined by the \compositionDim{panel uniformity} and the \compositionDim{connection structure} (\figureRef{fig:composition}).
It determines the possible ways the device can fold and unfold, as well as the three-dimensional shapes it can form.
\compositionDim{Panel uniformity} describes whether all panels of a device are identical in \compositionDim{size} and \compositionDim{shape} or whether they differ.
The \compositionDim{connection structure} defines the way the panels of a foldable device are connected to each other.
The topology can be linear, non-linear, or branched (see \figureRef{fig:composition} rows).
In linear topologies, panels are arranged along a straight line, and hinges are typically parallel to each other.
Hinges in non-linear cases appear in different orientations, and the panel arrangement can be irregular.
Branched topologies further extend this by allowing single hinges to connect multiple panels simultaneously (see \compositionDim{hinge connection cardinality}).

\subsection{Foldable Usage}
\label{subsec:usage-properties}
Beyond their composition, foldable devices are also defined by how they are used---their input methods, fold states, and usage modes.

\paragraph{Input Methods}
Just like today's regular mobile devices, foldables enable \usageDim{on-surface} and \usageDim{spatial} input, as well as \usageDim{folding}.
\usageDim{On-surface} input includes multi-touch and pen interaction on the device's display(s).
Multiple panels and varying \compositionDim{display distribution} enable simultaneous input across surfaces.
While potentially angled panels make multi-touch design more challenging~\cite{Li:TouchGestures:2022,hennecke2012,Can:AngleCAD:2022}, foldables open up new possibilities such as back-of-device gestures~\cite{shimon2015,cui2021}.
\usageDim{Spatial} input captures physical motion as commands, either through device movement or gestures like shaking or tilting (\eg via a built-in inertial measurement unit).
Physical folding can also serve as input when the angle between panels is sensed (\eg \cite{Bueschel:Foldable3D:2016,yeh2024}).
Three \usageDim{folding} input strategies are frequently used:
    (i)~folds as spatial states (threshold-based) that persist until a new geometric configuration is detected,
    (ii)~folding as continuous input, and
    (iii)~compound folding movements as gestures.
Fold interactions can also be paired with other input methods, such as touch, to realize clutching mechanisms (cf. \cite{hinckley2000,jones2012}).
In addition to these three main input modalities, foldables might still be combined with external input devices (\eg keyboard) and support interaction through speech or gaze.

\paragraph{Fold State}
We categorize fold states into \usageDim{planar} and \usageDim{non-planar}.
\figureRef{fig:states} provides examples of planar and non-planar states, including their sub-states, for varying panel \compositionDim{uniformity}.
 
(i)~In a \usageDim{planar} state, all panels lie within a single geometric plane.
This state spans a spectrum from minimal footprint, where panels are folded onto one another for portability (the closed state), to maximum footprint, where the device is fully unfolded to maximize visible display area.
Between these two extremes, intermediate planar states occur when a device remains flat but is only partially unfolded.

(ii)~In a \usageDim{non-planar} state, panels are folded out of a single plane, producing three-dimensional configurations, ranging from angled to volumetric.
Angled states occur when panels are folded at non-flat angles, forming configurations such as laptop, tent, or zig-zag folds.
Volumetric states arise when panels form (partially) enclosed volumes, such as triangular prisms, cubes, or other polygonal enclosures.
Our definition of volumetric states is comparable to what Gomes and Vertegaal~\cite{Gomes:PaperFold:2015} call 3D Shapes.
The boundary between angled and volumetric states can be ambiguous; some concave shapes could be classified as either.
\etal{Marquardt}~\cite{Marquardt:SurfaceConstellations:2018} propose similar categories (\emph{flat}, \emph{concave}, \emph{convex}, and \emph{closed}) for multi-monitor workspaces.
Their \emph{closed}, however, denotes enclosed shapes, not our minimal footprint state.

The \compositionDim{panel uniformity} significantly influences the attainable fold states, as variations may lead to asymmetrical or offset configurations.
For instance, \etal{Khalilbeigi}~\cite{Khalilbeigi:FoldMe:2012} distinguish between ``Centerfold'' and ``Partial fold'' layouts in two-panel devices, where partial folds use non-uniform panels and result in unequal closure.
Certain states, such as cubes, may require square \compositionDim{panel shapes} for clean alignment.
Also, as the \compositionDim{number of panels} and fold states increase, so does the complexity of content distribution, fold coordination, and interaction design.
Additionally, the device's overall shape and size change significantly during state transitions, greatly affecting displayed content.

\begin{figure*}
    \centering
    \includegraphics[width=\linewidth]{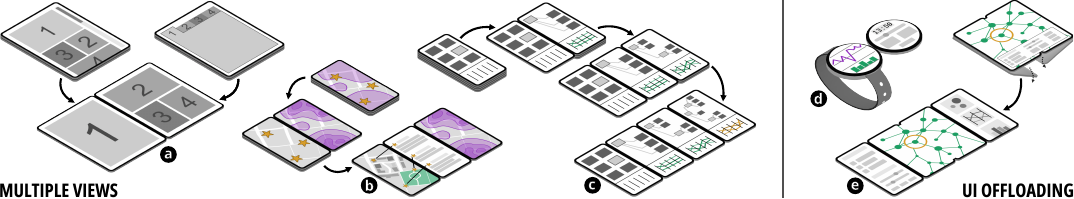}
    \caption{
    Foldables for \emph{multiple views} (left) and \emph{UI offloading} (right): 
    \figureLabel{a} Different view distribution/layout;
    \figureLabel{b} Switch design patterns of composite visualization views  (superimposition, juxtaposition, integration; cf.~\cite{javed2012});
    \figureLabel{c} Unfold panels to reflect stepwise analysis workflow/process;
    \figureLabel{d} Show a menu aside a visualization on a smartwatch;
    \figureLabel{e} Offload filters, alternative views, or controls for tools (\eg lenses).
    }
    \label{fig:views-offloading}
\end{figure*}

\paragraph{Usage Mode}
We distinguish three categories of the device's general physical disposition during use.
(i)~\usageDim{Handheld:}
The device is held in one or both hands, typically suited for casual information needs (i.e., mobile and glanceable~\cite{Blascheck:Glanceable:2019}) and influenced by various grip postures~\cite{Perelman:HingeGestures:2026}.
(ii)~\usageDim{Surface-supported:}
The device rests on a surface for support, including flat and non-planar states (\figureRef{fig:states}).
Multiple display surfaces visible from various angles (\eg \cite{Saniee-monfared:TentMode:2020}) can be exposed, supporting prolonged interactions and shared use.
(iii)~\usageDim{Wearable:}
Some configurations might allow the device to be carried in compact or body-attached forms~\cite{Fuchs:FoldWatch:2018}.
These smaller formats prioritize portability.
Similar to semi-fixed device setups~\cite{Marquardt:SurfaceConstellations:2018}, the usage mode is dynamic.
It is based on the fold state and interaction context, and constrains how many display surfaces are visible at once and which input methods are practical.
\section{Visualization Concepts for Foldables}
\label{sec:vis-concepts}

Building on our device-centered design space, this section presents a conceptual framework that specifically explains and showcases how visualizations can adapt to and leverage foldables.
Our goal for this framework was to integrate established task taxonomies~\cite{amar2005,yi2007,brehmer2013} and various visualization techniques, data types, and usage scenarios.
We targeted concepts that, in our view, benefit from foldable form factors.
As part of the creative process, the team of authors engaged in both an intense two-week in-person meeting phase and several remote discussions.
We generated ideas, detailed them using wireframes and paper prototypes, clustered them into groups using affinity diagramming, and reflected on the coverage of data analysis goals and interactions, which helped refine and enrich the framework iteratively.
In parallel, our low-fidelity ideation workshop (\ref{sec:prototyping:workshop}) drew on initial concepts and fed ideas back into our discussion.
The author team comprises three senior researchers and one junior researcher with backgrounds in visualization and HCI (two professors and one postdoc, each with $15$+ years of research experience, and one second-year PhD student).

This section outlines the concepts that resulted from this process, organized along four complementary perspectives:
    (\ref{sec:more-displays})~\textbf{more displays},
    (\ref{sec:more-shapes})~\textbf{more shapes},
    (\ref{sec:more-interactions})~\textbf{more interactions},
    (\ref{sec:more-states})~\textbf{more states}.
We chose these to illustrate various ways to leverage the unique characteristics of foldables for visualization.
At the same time, we acknowledge that the individual perspectives overlap.

\subsection{\textbf{More Displays}: Distributing Visualization Views}
\label{sec:more-displays}

In its most basic form, a mobile device capable of unfolding allows for enlarging a view. 
This is related to responsive visualization~\cite{Horak:Book-Responsive:2021,hoffswell2020}, as views can adapt to changes in screen space in various ways (\eg level of detail, arrangement of visual elements, visual encoding).
Beyond this, the extended display space can be used effectively to accommodate additional views (a key principle in many visualization applications~\cite{roberts1998encouraging,Chen2021CompositionPatterns}), or other essential user interface (UI) components.

\subsubsection{Multiple Views}
The limited screen space of conventional mobile devices restricts the use of multiple coordinated views, where different encodings, juxtaposed and interconnected via linked brushing, support different tasks~\cite{roberts1998encouraging}.
With their expandable multi-panel design, foldables offer several ways to address this constraint.
\examples{}
When unfolded, they may provide enough screen space to distribute coordinated views across panels, thus supporting dynamic grid layouts beyond scrollable or tabbed interfaces (\figureRef[a]{fig:views-offloading}).
In addition to view layout, (un)folding can be used to switch between design patterns of composite visualization views~\cite{javed2012}.
A superimposed view on a single panel can be split into separate side-by-side layers (\figureRef[b]{fig:views-offloading}), similar to the \emph{foldable layers} by \etal{Khalilbeigi}~\cite{Khalilbeigi:FoldMe:2012}.
Furthermore, unfolding actions seem to suit stepwise exploration processes.
Assigning views on demand on sequentially unfolded panels (\figureRef[c]{fig:views-offloading}) could mirror the temporal nature of analytic workflows and facilitate comparisons across variations of views (\eg \cite{Korn:PMC-VIS:2023}).
\sectionEnd{
By distributing views across panels and linking layout changes to folding actions, foldables address both the spatial arrangement of multiple views and the stepwise nature of analytic workflows.
They can therefore function as reconfigurable analytic workspaces, where the physical configuration shapes how users arrange and access views.}

\subsubsection{UI Offloading}
In addition to the actual data items that are displayed, visualization applications usually include additional components such as legends, descriptive texts, tooltips, and details on demand, filter options, or settings dialogues.
Similar to other cross-device interfaces (\eg \cite{Langner:VisTiles:2018, Langner:Marvis:2021}), we believe that additional unfolded panels are ideal for displaying these UI components separately, but still close to the actual data.
\examples{}
On a foldable smartwatch, for instance, certain UI elements can be made visible by unfolding the device (\figureRef[d]{fig:views-offloading}).
This principle extends to devices with more panels and to further interface components.
\figureRef[e]{fig:views-offloading} shows a gate fold device.
To allocate more space for the node-link diagram, on which a circular magic lens~\cite{Tominski:Lens-Survey:2017} operates, the left panel displays parameters of the lens, while the right shows alternative visualization techniques.
As the number of panels grows, so do the possibilities for distributing these components, for example, dedicating individual panels to specific functions like filtering, configuration, or alternative views.
\sectionEnd{
These examples suggest that moving UI components onto separate panels can reduce occlusion of the visualization and support more structured layouts.
Dedicating panels to UI elements this way may also widen the range of possible interactions.}

\subsubsection{Detail-and-Context Techniques}
Overview+detail and focus+context are established techniques for navigating large information spaces~\cite{cockburn2009}; we use \emph{detail-and-context} as an umbrella term for both.
On mobile devices, limited screen space constrains both: overview and detail views compete for space~\cite{goncalves2012,burigat2013}, while focus+context techniques struggle to balance readable focus with sufficient context.
Foldables can address this by distributing detail and context views across separate panels.
\examples{}
For detail-and-context, views distributed across panels are semantically connected, relating through meaning or function, as illustrated in \figureRef[a]{fig:detail-context-shapes}, where a geographic map provides an overview of objects, and a list of corresponding time series is displayed.
Views can additionally be spatially connected, relying on the physical continuation across hinges to form a coherent visual layout.
For example, in \figureRef[a]{fig:detail-context-shapes} an overview line graph (\eg stock quotes) spans the upper panel and a magnified detail view is aligned beneath it, with visual elements highlighting this relation.
For devices with more than two panels, multiple detail views can support parallel exploration of different locations (\figureRef[a]{fig:detail-context-shapes}), avoiding the need to sequentially pan a single detail view between locations~\cite{saidi2016}.
\figureRef[b]{fig:detail-context-shapes} illustrates that even non-rectangular panels, such as a triangular one, can serve as detail views, \eg to display an adjacency matrix highlighting relations between listed data items.
Focus+context on a foldable can, for example, be realized as a bifocal display~\cite{spence1982}, where one panel serves as the focus and adjacent panels provide spatially connected context (see \figureRef[c]{fig:detail-context-shapes}).
\sectionEnd{
By distributing detail and context views across panels, foldables extend these techniques to multi-panel devices.
For non-continuous display surfaces, however, this separation means that focus+context is no longer integrated into a single view, blurring the boundary with overview+detail.
Nonetheless, the spatial separation could also support user cognition by making the roles of panels explicit.}

\begin{figure*}
    \centering
    \includegraphics[width=\linewidth]{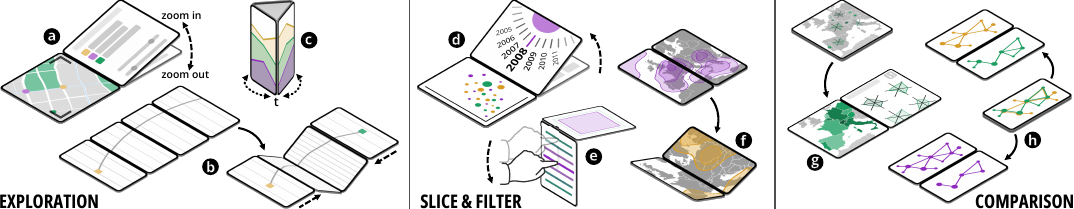}
    \caption{
        Foldables for \textit{exploration} (left), \textit{slicing \& filtering} (middle), and \textit{comparison} (right):
        \figureLabel{a} Continuous fold angle changes to control parameters (\eg zoom level);
        \figureLabel{b} Physical folding of an information space (compression);
        \figureLabel{c} Explore a temporal information space by rotating (left-right panning);
        \figureLabel{d} Select a filter category from a list/stack by fold angle;
        \figureLabel{e} Control a data slice on an orthogonal panel;
        \figureLabel{f} Navigate layered information space via panel angle;
        \figureLabel{g} Change comparison strategy (nested to juxtaposed);
        \figureLabel{h} Separate superimposed into juxtaposed views or using explicit encodings.
        }
    \label{fig:exploration-slice-filter}
\end{figure*}

\subsection{\textbf{More Shapes}: Utilizing Spatial Configurations}
\label{sec:more-shapes}
Foldables not only allow to change the size of the available screen area, but also its spatial arrangement.
In this part, we examine how visualization views can be mapped to the spatial configuration of connected panels and how common visualization techniques can leverage this.

\subsubsection{Structural Alignment (Axes, Layers, Hierarchies)}
\label{subsubsec:structural-alignment}

As described in \ref{subsec:composition}, display surfaces might not always be continuous across panels.
Hinges and folding creases may disrupt the perception of continuity, similar to findings in research on tiled display walls where bezels interrupt visual flow \cite{dealmeida2012,wallace2014}.
Mapping a visualization across all panels of a foldable can therefore lead to perceptual issues.
However, some visualization techniques can directly make use of these breaks, such as positioning axes at their location so data items are not obscured.
\examples{}
A simple example is placing two charts on adjacent panels with a shared axis.
\etal{Deng}~\cite{deng2023} describe this symmetric arrangement of visualizations as a mirror pattern.
It invites comparison and is visually appealing because of its symmetry, but it is limited to two charts and makes small differences harder to see compared to overlays or explicit difference encodings.
Visualization techniques that feature multiple, parallel axes can also make use of this idea.
Take \figureRef[d]{fig:detail-context-shapes} for example, where the hinges and the outer panels' edges are used to place the axes of a parallel coordinate plot.
Here, folding can support simple analytic actions, such as hiding dimensions or bringing non-adjacent axes together for comparison.
Similarly, hierarchical structures can map depth to the device's composition.
For example, see \figureRef[e]{fig:detail-context-shapes}, where the levels of a Degree-of-Interest Tree~\cite{card2002} can be distributed across panels (one hierarchical level per panel).
Here, the hinge acts as a delimiter between parent and child nodes.
This physically enforces the logical hierarchy.
\sectionEnd{Aligning visualization structures with the physical seams of a foldable turns potential visual disruptions into delimiters that reinforce the data's organization.
However, the number of simultaneously visible panels limits how many structural elements can be displayed at once.}

\subsubsection{Spatialized Views}
Foldable devices can assume spatial configurations that extend beyond a single planar surface.
By distributing content across angled or volumetric states (see \figureRef{fig:states}), foldables allow visualization views to be arranged in three-dimensional space, where the device's geometry itself conveys spatial relationships between views.
This can benefit techniques that simulate folding, whose visual structure naturally maps to a specific fold state, or that represent inherently spatial data.
\examples{}
Some of these complete shapes benefit from a surface-supported setting.
Take, for example, a linear three-panel foldable with two angled panels, as shown in \figureRef[f]{fig:detail-context-shapes}.
Here, the central panel can display an undistorted focus view, while the side panels fold back and show contextual information, resembling the perspective wall~\cite{mackinlay1991} technique.
Foldables can also reproduce actual three-dimensional visualizations.
For example, in \figureRef[g]{fig:detail-context-shapes}, a laptop-like state is used to display spatiotemporal data.
The lower panel shows a map with selected entities, while the angled panel presents how the data evolves over time along that selection.
This resembles the space-time wall~\cite{Tominski:Wall:2012}.
Compositions with square panels that can be folded into cube-like states (at least 3 orthogonal panels) are particularly promising.
Such shapes lend themselves to pseudo-3D visualization, as each panel can host a projection of a higher-dimensional dataset.
In \figureRef[h]{fig:detail-context-shapes}, three corresponding attribute combinations of a scatterplot matrix can be folded up into this semi-cube structure.
A fourth panel could be used to navigate the scatterplot matrix.
A (semi-)cube fold with outside displays could enable tangible implementations of 3D navigation techniques such as ScatterDice~\cite{Elmqvist-rolling-the-dice-2008}, and suits techniques like the hive plot shown in \figureRef[i]{fig:detail-context-shapes}.
With head-coupled perspective tracking, such configurations can present data as if it were located inside the device, akin to fish tank VR~\cite{ware1993}, but achieved through physical panel arrangement.
\sectionEnd{
Unlike flat displays, foldable devices can leverage their physical dimensionality as part of the visual encoding, making the device's geometry itself an active element of the visualization.}

\subsection{\textbf{More Interactions}: Folding for Visualization Tasks}
\label{sec:more-interactions}

While the previous concepts mainly addressed how visualizations are distributed and displayed, this perspective focuses on their interaction, with a particular emphasis on folding.
Additionally, we consider spatial interaction enabled by specific fold states.
The following concepts are derived from and inspired by visualization tasks and activities~\cite{amar2005,yi2007,brehmer2013} as well as composite visualizations~\cite{javed2012,deng2023}.

\begin{figure*}
    \centering
    \includegraphics[width=\linewidth]{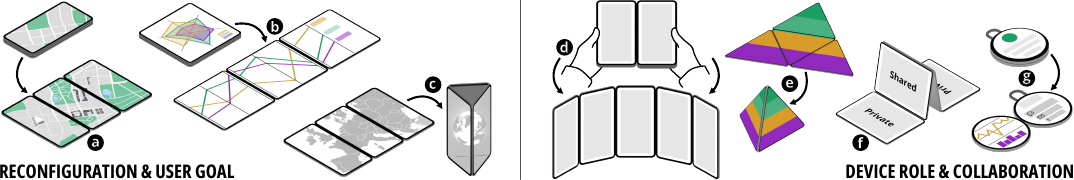}
    \caption{
        Foldables for visualization \emph{reconfiguration \& user goals} (left) and \emph{device roles \& collaboration} (right):
        \figureLabel{a} Using (un)folding as semantic zoom to access different levels of detail;
        \figureLabel{b} Changing visual encoding from a star plot to a parallel coordinates plot;
        \figureLabel{c} Turning a 2D map (flat state) into a 3D globe (volumetric state);
        \figureLabel{d} Moving from handheld to surface-supported use;
        \figureLabel{e} Using a volumetric state to support view visibility and readability for collaboration;
        \figureLabel{f} Combining shared and private views in laptop-like states;
        \figureLabel{g} Separating public from private data on a foldable locket.
    }
    \label{fig:more-states}
\end{figure*}

\subsubsection{Exploration}
When screen space is limited, only a subset of a dataset can typically be displayed.
Here, we focus on how folding and fold states can be used to explore these subsets, following \etal{Yi}'s~\cite{yi2007} notion of exploration.
\examples{}
In its simplest form, changing the angle between two panels can be mapped to a specific value.
In these cases, one panel could display the visualization while a separate panel is used for interaction.
For example, in \figureRef[a]{fig:exploration-slice-filter}, the angle between two panels controls the zoom factor of a map visualization.
Distortion-based exploration methods can fold or compress an information space to bring distant regions closer together.
On foldable devices, this folding can be physicalized.
One such example, shown in \figureRef[b]{fig:exploration-slice-filter}, is the M\'elange distortion technique~\cite{elmqvist2010}, where folding the middle panels of a linear four-panel device compresses the space between two points of interest.
Spatial interaction can also emerge from specific fold states.
For example, a volumetric state (\figureRef[c]{fig:exploration-slice-filter}) can display time series data wrapped around the device, where users navigate by rotating.
Upon completing a full rotation, the displayed subset can update, allowing the interaction to continue across multiple rotations.
This could extend to other visualizations with ordered dimensions, such as parallel coordinate plots.
\sectionEnd{These examples illustrate that foldable devices can offer exploration mechanisms that go beyond touch input, using folding and fold states.}

\subsubsection{Slice \& Filter}
Through techniques such as slicing, filtering, or layering, subsets of variables or dimensions can be presented at a time.
\examples{}
Instead of zooming, a filter parameter could also be controlled through folding.
As an example, \figureRef[d]{fig:exploration-slice-filter} shows an adaptation of Gapminder's Bubbles tool~\cite{gapminder_tools}, where the folding angle controls the displayed time step.
Clutching a specific activation area could select the parameter to filter, with the fold angle controlling the filter value.
As suggested by previous work~\cite{Bueschel:Foldable3D:2016,Can:AngleCAD:2022}, angled panels can enable touch interaction across three dimensions.
For example, panels in states such as in \figureRef[e]{fig:exploration-slice-filter} can be used to control a quasi-orthogonal clipping plane, similar to spatial interaction to slice through volumes~\cite{Steimle:Flexpad:2013,Spindler:PaperLens:2009}.
The same principle can navigate layered information spaces, analogous to how PaperLens~\cite{Spindler:PaperLens:2009} lets users move through stacked layers.
In both cases, touch input on the angled panel substitutes for spatial movement of the device.
Many analytic tasks require coordinating subtasks like navigation, selection, and filtering within a single interaction phrase~\cite{buxton1995}.
In \figureRef[f]{fig:exploration-slice-filter}, touch input with both thumbs controls zooming or panning of a map, while the angle between the panels controls the active data layer (cf.\;\cite{kildal2013}).
This lets users interleave or perform subtasks simultaneously, a form of bi-manual, multimodal interaction.
\sectionEnd{%
Foldable devices support multi-modal interaction (slice \& filter) that combines touch and fold input.
It is apparent that with layered information spaces, foldables enable both direct (select a slice, \figureRef[e]{fig:exploration-slice-filter}) and indirect (angle mapped to layered space, \figureRef[f]{fig:exploration-slice-filter}) interaction.}

\subsubsection{Comparison}
Comparison is central to many visualization tasks, with juxtaposition, superposition, and explicit encoding being common strategies~\cite{gleicher2011,lyi2021}.
We believe foldable devices can extend these approaches by allowing users to transition between strategies through folding.
\examples{}
\figureRef[g]{fig:exploration-slice-filter} illustrates that unfolding can transform nested, overloaded, or similar views into juxtaposed layouts to support different comparison goals.
In a closed state, nested or overlapping views support comparison by spatial arrangement, \eg neighboring countries on a map.
Unfolding decomposes these into juxtaposed arrangements, enabling attribute-based comparison, ordering, and filtering.
For superimposed views, the user could use specific touch areas or clutching to choose which comparison pattern to display upon unfolding.
For example, in \figureRef[h]{fig:exploration-slice-filter}, unfolding can produce either a simple juxtaposition of the two views or an explicit juxtaposition, where one panel shows the union and the other shows the difference.
\sectionEnd{%
Foldables provide a physical, reversible transition between established comparison strategies: unfolding separates visual representations, and folding merges them.
When the initial state is a combined layout (\eg superimposed, nested, or overloaded), unfolding separates the panels, thereby separating the representations of data objects.
Conversely, folding panels together merges juxtaposed elements.}

\subsection{\textbf{More States}: Employing Responsive Visualization}
\label{sec:more-states}

Foldables allow users to switch between compact and expanded states.
This changes both the available display area and the spatial arrangement, as well as how the device can be held, positioned, or shared.
In the context of responsive visualization~\cite{Horak:Book-Responsive:2021}, these fold-state transitions serve as \emph{physical breakpoints}, complementing software-defined breakpoints~\cite{Schottler:Breakpoints:2025}.
This section describes how such transitions can affect visualization configuration, user goals, and how certain fold states change a device's role and collaboration possibilities.

\subsubsection{Reconfiguration \& User Goal}
\label{subsubsec:reconfiguration}
The most immediate response to a fold-state change is to adapt the visualization to the new display configuration.
\examples{}
At the simplest level, (un)folding can rearrange visual elements to fit the current display geometry, such as switching the orientation of a bar chart~\cite{hoffswell2020,Horak:Book-Responsive:2021}.
Beyond this, folding can also control the level of detail.
A semantic zoom technique could progressively reveal more detailed content with each unfolding action, such as roads and buildings on a map (\figureRef[a]{fig:more-states}).
Another aspect of responsive visualization is changing user goals or usage context.
A foldable smartwatch could provide a glanceable visualization~\cite{Blascheck:Glanceable:2019} during a workout, and allow a detailed exploration with unfolded additional panels at home (similar to \figureRef[d]{fig:views-offloading}).
This might go hand in hand with a shift regarding the \emph{intended viewing timespan} and \emph{complexity of visualization interaction}~\cite{langner:book-intro:2021}.
Folding can also induce changes between visualization techniques altogether (i.e., responsive changes to visual encoding~\cite{Horak:Book-Responsive:2021}).
\figureRef[b]{fig:more-states} shows such a transition by changing a star plot to a parallel coordinates plot.
Further, this could be a change from a more compact adjacency matrix to a more widespread node-link diagram~\cite{gladisch2015}.
Where applicable, foldables can enable additional transitions driven by volumetric fold states, \eg switching a 2D map to a 3D globe (\figureRef[c]{fig:more-states}).
This follows previous design recommendations~\cite{Gomes:PaperFold:2015}.
\sectionEnd{
Folding acts as a trigger for responsive visualization.
From rearranging visual elements and adjusting detail to switching visualization techniques in response to changing device and usage factors.
}

\subsubsection{Device Role \& Collaboration}
Beyond adapting what is shown, fold-state transitions also change how the device is used and shared, adding device role and collaboration as a further dimension of responsive behavior on foldables.
\examples{}
While foldables are primarily handheld, certain configurations may require surface placement due to size or complexity.
\figureRef[d]{fig:more-states} shows this transition from handheld to surface-supported use.
Once placed on a surface, a device's fold state dictates its collaborative potential. 
Unlike flat tablets, which may impose awkward viewing angles, foldables can be arranged to orient panels toward different users. 
For example, a tetrahedral configuration with four panels can support three distinct viewing directions (\figureRef[e]{fig:more-states}), while other folds enable face-to-face collaboration (\figureRef[f]{fig:more-states}).
Here, flat-lying panels can provide private views and upright sections serve as shared workspaces, mirroring the physical changes between individual and collaborative workspace explored in KirigamiTable~\cite{Gronbaek:KirigamiTable:2020}.
This idea extends to managing information visibility on more personal devices.
A foldable worn as a locket could display outward-facing data (\eg mood or group affiliation) when closed and personal details when opened (\figureRef[g]{fig:more-states}).
\sectionEnd{
Foldables are unique in that a physical reconfiguration can transition a device between individual and collaborative visualization use, from handheld to surface-supported, and from private to shared views.
}
\section{Prototyping Visualizations for Foldables}%
\label{sec:prototyping}
To move from the theoretical design space and conceptual framework towards practical experience, we engaged in prototyping activities.
This section describes an ideation workshop we conducted (\ref{sec:prototyping:workshop}) as well as our strategies for creating interactive visualization artifacts (\ref{sec:prototyping:artifacts}).

\subsection{Low-fidelity Ideation Workshop}%
\label{sec:prototyping:workshop}
We conducted an envisioning and ideation workshop to explore how the identified design dimensions for foldable devices can be translated into concrete visualization concepts and usage scenarios.

\subsubsection{Method and Design of the Workshop}
The workshop combined paper prototyping and sketching.
The low-fidelity approach aimed to foster creativity and reduce the risk of being overly constrained by technical considerations.

In accordance with local ethics regulations, this type of envisioning activity is exempt from a formal ethics committee review.
We followed standard ethical guidelines: all participants provided written informed consent (see supplemental material), took part voluntarily, could withdraw at any time without consequences, received fair compensation, and are attributed anonymously.

\smallParagraph{Participants}
We recruited 15 computer science students (8 male, 6 female, 1 undisclosed) aged 21 to 30 years ($M = 25.1$, $SD = 2.0$) through the mailing list of a class on Information Visualization.
Twelve were enrolled in a master's program, 2 bachelor's, and 1 undisclosed.
There were no other selection criteria.
On a 5-point rating scale, 2 participants rated their visualization experience as novice, 1 beginner, 9 intermediate, and 2 advanced (no experts, 1 undisclosed).
No participant owned a foldable.
Participation was compensated with 20 EUR, unrelated to any course assessment and, as mentioned above, voluntary.

\smallParagraph{Procedure}
We ran two sessions of the workshop, and participants were assigned to 5 groups, each with 2-4 people.
Each session lasted 90 minutes and was supervised by three instructors.
After obtaining written informed consent from all participants, explaining the general topic, and showing some inspirational images of foldable mobile devices and paper-folding techniques, the groups developed ideas ($\sim$30 min.) and presented them to one instructor.
In a second step, the groups refined their designs with a focus on possible interaction techniques ($\sim$30 min.) and concluded with a short presentation.
Each group was provided with
    crafting materials,
    pre-cut cardboard shapes (square, rectangular in a 1:2 ratio, and round) in three sizes (similar to wearable, smartphone, and tablet),
    structured sketching sheets (inspired by the FDS methodology~\cite{roberts2016}),
    a tablet showing a visualization catalog website~\cite{datavizproject}, and
    a printed list of common visualization interactions~\cite{yi2007} and low-level analytical tasks~\cite{amar2005}.
At any time, participants were free to combine or modify the cutouts or create entirely new shapes from paper.

\begin{figure}[t]
    \centering
    \includegraphics[width=\linewidth]{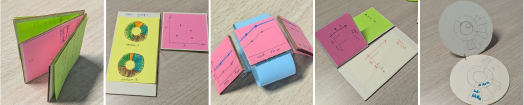}
    \caption{
        Five exemplary results of our ideation workshop in the form of paper prototypes
        (from left to right: \idea{I1}, \idea{I9}, \idea{I11}, \idea{I13}, and \idea{I18}).
    }
    \label{fig:workshop}
\end{figure}

\smallParagraph{Data Analysis}
We video-recorded the presentations (transcribed with \href{https://otter.ai}{otter.ai}) and collected the resulting paper prototypes, each with a sketching sheet that provided explanations and additional sketches.
Two authors independently coded the data and reached consensus on themes and findings through multiple discussions.
We summarize the findings below and provide a detailed table in the supplementary material. 

\subsubsection{Workshop Outcomes}
The participants created 19 distinct prototype ideas (\idea{I1}--\idea{I19}) in total (\figureRef{fig:workshop} shows five examples).
The utilized form factors and visualization tasks cover many of the design dimensions described in \autoref{sec:foldable-design-space} and the concepts in \autoref{sec:vis-concepts}.  
Note that we specify only one prototype idea in each case here, though most were present across multiple ideas.

\smallParagraph{Use Cases and Form Factor}
Nine out of 19 ideas are for health monitoring or activity tracking using small, portable devices and wearables.
The remaining ideas include
    entity comparison (\eg products or students),
    trend analysis with aggregated and detailed data (\eg stocks), and
    geological data on maps.
Smaller form factors were the most popular:
    9 prototypes are handheld with a size mostly smaller than commercial foldable phones, and 8 even smaller prototypes are body-worn. 
Only in 4 prototypes was a fold state intended for surface support considered.
The participants constructed foldables with
    diverse panel shapes (rectangle \idea{I5}, triangle \idea{I6}, circle \idea{I7}, and cut-off circle \idea{I4}),
    various connection structures (linear \idea{I15}, non-linear \idea{I6}, branched \idea{I9}), and
    different hinges (line \idea{I2}, branched \idea{I1}, point joint \idea{I18}).
The designs included single-sided (\idea{I3}) and double-sided (\idea{I4}) displays while combining both uniform (\idea{I1}) and non-uniform panels (\idea{I5}).

\smallParagraph{View and UI Distribution} 
Participants assigned views and UI controls to panels both statically and dynamically.
In static configurations, panels are dedicated to fixed roles, \eg
    two panels reserved for patient information in an emergency care booklet (\idea{I1}) and
    two side panels for UI controls inspired by game controllers (\idea{I5}).
Views also show different encodings of the same data (\idea{I4}) or different entities for comparison (\idea{I18}).
Other designs make view configurations responsive to changes in fold state.
For example, a cube display could serve as a passive overview when positioned like a piece of furniture, unfold into a docked wide display for important information or task reminders, and extend fully into a larger flat display to analyze complex data (\idea{I2}).

\smallParagraph{Visualization Design and Key Interaction}
In terms of visualization, 8 prototypes use an overview+detail technique.
In most of these cases, folded states provide overview only, and unfolded states show both an overview and a detail view.
Two ideas explicitly involve focus+context, such as a fisheye lens (\idea{I4}) or a perspective wall-inspired technique (\idea{I17}).
The ideas also cover different data types, most often multivariate (\idea{I15}), but also temporal (\idea{I12}), spatial (\idea{I14}), and layered (\idea{I16}).
Moreover, the participants designed different interactions.
For example,
    changing the fold state to trigger view reconfiguration (\idea{I3}),
    using the fold angle to determine whether two panels are considered separate or joined to form a single large view (\idea{I5}), and
    folding two panels together to aggregate the data shown on them.
A continuous folding movement was suggested for zooming (\idea{I2}) or selecting data layers (\idea{I16}).
In one case (\idea{I18}), a point hinge was installed as a means of increasing the degree of freedom in folding, effectively providing an additional axis of movement to, \eg navigate through a list.
Besides folding, most ideas incorporated touch input.
For free-standing surface-supported devices, participants suggested mid-air hand gestures and gaze input (\idea{I2}).

\subsection{Interactive Prototypes}
\label{sec:prototyping:artifacts}

\begin{figure}[t]
    \centering
    \includegraphics[width=\linewidth]{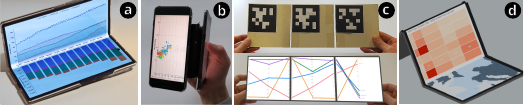}
    \caption{Our four strategies of interactive prototyping:
        (a) commercial foldables,
        (b) combining regular smartphones,
        (c) MR-overlays on cardboard prototypes, and
        (d) rendered animations.}
    \label{fig:prototypes}
\end{figure}

To get a more tangible impression of our concepts than paper prototypes can provide, we developed interactive visualization artifacts for foldable form factors.
However, no single prototyping approach currently achieves physical realism, full interaction support, and flexible form factors simultaneously.
We therefore experimented with four strategies that illustrate visualization examples from our conceptual framework, each setting different priorities.
The actual interactive visualizations were implemented as web-based applications for all strategies.

First, we develop on a \textbf{commercially available foldable} (Samsung Galaxy Z Fold 7) to prototype visualizations with chart elements aligned across panels (see \ref{subsubsec:structural-alignment}; \figureRef[a]{fig:prototypes}).
This approach provides the highest physical realism and good interaction fidelity but is constrained to two or three rectangular panels, no \ang{360} folding, and limited API support (\eg the W3C \textit{Device Posture API}) for accessing fold angles.

Second, we combine \textbf{regular smartphones} with 3D-printed hinges (\figureRef[b]{fig:prototypes}) to prototype continuous fold-angle input for data exploration.
Our implementation replicates the Gapminder bubble chart example to browse yearly economic world data on two joined Honor~9 devices.
Using inertial sensors, we realized two modes:
    (i)\,when unfolded, touch input triggers a sliding time window (cf.\,rapid serial visual representation, RSVP), whose speed is controlled by the folding angle; and
    (ii)\,when perpendicular, data is shown on the top panel and years are navigated via touch-drag on the side panel.
While this strategy exploits the display and touch input capabilities of modern mobile devices and adds fold-angle sensing, it is challenging to integrate numerous panels and combine displays of different shapes and sizes.

Third, we prototype folding-driven responsive visualizations (see \ref{subsubsec:reconfiguration}) on devices with more than two panels.
We attach ArUco markers to cardboard panels and render visualizations on them using \textbf{mixed reality glasses} (Meta Quest 3).
Based on a dataset of football player statistics, we implemented the transition from a star plot in the closed state to a parallel coordinates plot in the unfolded state (\figureRef[c]{fig:prototypes}).
However, this strategy is not ideal for touch interaction.
The hand tracking accuracy in such a setup can be very limited (depends on hardware), and hands may cover the markers.

Fourth, we create \textbf{virtual video prototypes} to envision and communicate visualization concepts for arbitrary foldable form factors, using a 3D animation tool.
We map screencasts of visualization views onto an animated 3D model of a three-panel device.
In a flat state, the device shows a choropleth map of European unemployment rates.
In a laptop-like state, the vertical panel displays a heatmap of countries selectable on the bottom panel (\figureRef[d]{fig:prototypes}).
Folding the panels into a triangular prism, the same data appears as a 3D globe floating inside the device.
This approach maximizes flexibility in form factor and visual presentation, but the artifacts are non-interactive and not tangible.

We found each strategy valuable for different purposes.
Commercial foldables and combined smartphones provide the most realistic interaction;
    mixed reality offers great form-factor flexibility and, given stable hand/object tracking, can be powerful;
    while virtual video prototypes are suited for communicating early concepts (\eg to stakeholders).
\section{Discussion}
\label{sec:discussion}

We reflect on limitations, implications, and further opportunities of our work about foldables for visualization---that is our device-centered design space, conceptual framework, and prototyping efforts.
With this, we also aim to promote future VIS/HCI research on the topic.

\paragraph{Limitations and Scope of this Work}
Our contributions are design-driven, and several choices define the scope of our exploratory work.
First, our design space (\autoref{sec:foldable-design-space}) targets handheld devices with flat, non-detachable, foldable panels.
We set aside rollable, stretchable, and other continuously deformable or actuated devices keeping the space grounded in widely adopted screen products.
Second, our framework (\autoref{sec:vis-concepts}) is tailored to data visualization.
Yet foldables are versatile, and the general mechanics of some of our concepts (\eg multiple views, offloading UI components, filtering, device roles) potentially transfer to other interactive applications.
Third, we intend the four perspectives as a lens for organizing concepts and not as a complete or mutually exclusive classification.
The perspectives overlap, and realistic use cases combine them, which reflects how the concepts are meant to be used together.
We drew on selected visualization task taxonomies~\cite{amar2005,yi2007,brehmer2013} to generate and structure concepts, while task-specific guidance and best practices remain future work.
Fourth, our ideation workshop (\autoref{sec:prototyping:workshop}) was not designed to evaluate usability or effectiveness but to broaden the concept space.
Its 15 participants, with mostly intermediate experience, offer an early perspective on the possibilities we describe.

\paragraph{Form Factor}
Current foldables primarily target the phone-to-tablet transition.
Our concepts deliberately include compositions beyond this (non-rectangular panel shapes, non-linear connections).
Participants in our workshop did not hesitate to use non-rectangular panels.
They generally favored smaller panels, suggesting a preference for compact, portable devices.
The latter, in our view, is because smaller devices can benefit much more from additional screen space in the unfolded state.
We are convinced that, alongside the typical smartphone form factor, particularly small devices (\eg wearables, jewelry) should take advantage of revealing displays by unfolding.
\textit{Consequently, foldables should be explored not only as generic, one-size-fits-all devices, but also for scenarios that call for specialized devices with different form factors (in the sense of `data follows form' or `form follows data').}

\paragraph{Prototyping}
Our prototyping experiments show that even simple setups can help to make foldable visualization concepts tangible.
Prototyping in this field is challenging; there exist only few commercially available devices.
Also, building your own innovative folding devices can quickly become technically complex.
For instance, the ideation workshop led participants to envision designs such as circular panels with a point hinge (\idea{I18}) that would be difficult to physically realize.
That complexity also goes beyond hardware.
Simulated design environments for foldables (\eg \cite{khaddam2020}, Android Studio Emulator) are an option, as these can help develop and illustrate ideas.
\textit{However, there is a real need for research toolkits that cover not only general user interface aspects but also accommodate visualization-specific requirements, thereby allowing for hands-on, tangible experimentation.}

\paragraph{Device Handling}
We assume that the hinges of future foldables will allow for easy opening and closing without much exertion, as well as secure locking.
Therefore, we see two challenges regarding how users handle foldable devices.
First, the number of panels of a foldable may directly impact its usability.
For example, a user of a device with 5 or more linearly arranged panels could fold each panel individually and adjust the fold angles, or alternatively, it could be enough to only grab the outer panels and pull them apart or push them together, much like an accordion.
\textit{Hence, we suggest investigating ergonomic issues such as where and how to best grab and hold more complex foldables, and whether they should be placed in the hands, on the lap, or on a table.}

Second, foldables seem more prone to unintended input and accidental activation.
This is mainly due to (i)\,the described dynamics of grabbing and holding a foldable, which is particularly problematic when handling double-sided touchscreen panels; and
(ii)\,the distinction between folding as direct, continuous input (\eg \figureRef[a+d]{fig:exploration-slice-filter}) and folding to change a device's fold state (\eg opening, closing).
\textit{To spare users the frustration associated with these uncertainties, we propose further research on, for example, explicit and implicit mode switching, clutching approaches, as well as detecting and capturing user intent.} 

\paragraph{Beyond Casual \& Personal Visualization}
Modern mobiles are primarily personal tools.
Mobile data visualizations often focus on use cases within personal or casual contexts.
The limited screen space typical of mobile devices further reinforces simplicity and glanceability (\eg \cite{Blascheck:Glanceable:2019}) of the content displayed.
However, foldables offer a more dynamic and adaptable screen arrangement and open up the potential to go beyond simple, casual, and personal visualization.
For example, when a user needs more screen space to move from quickly reading a value to conducting a deeper investigation into its causes and relationships, the device can simply be opened up.
This action not only allows for additional views or more detailed information, but could also unlock entirely new application functionalities and interaction possibilities.
\textit{For this reason, we invite future research that investigates this transition between personal and professional visualization, its seamlessness and continuity regarding both visual and interaction aspects (\eg animations, handheld vs. surface-supported as shown in \figureRef[d]{fig:more-states}).}

\paragraph{Folding Ambiguity}
Besides having to learn a set of folding techniques (interaction vocabulary, discoverability), our concepts reveal a level of ambiguity associated with folding.
One fold state cannot always be associated with exactly one visualization adaptation (\eg alternative layout, zoom, filter)---there are multiple ways a visualization can react to folding.
This also implies potential challenges for users in correctly anticipating the system's responses and the consequences of their interactions.
\textit{We therefore see the need to further develop and investigate solutions that clearly communicate to users what happens as a result of device folding (cf.\,feedforward), allow for user configuration, and lay the foundations for foldable visualization conventions.}
\section{Conclusion}
In this work, we explored the potential of novel foldable devices for data visualization.
Our design space exploration indicates that foldable compositions and their usage potential are vast, and we see great opportunities for foldable form factors to diversify further, \eg into wearables.
This was also supported by our envisioning workshop, where participants created diverse prototypes and generally favored smaller devices for personal use.
Equally, the potential of foldables for visualization is not limited to providing additional screen space.
As organized by our conceptual framework, view configurations can respond to fold-state changes, spatial device states can give physical form to data, and folding itself can serve as an input channel.
At the same time, both the workshop and our prototyping efforts exposed a gap between what can be envisioned and what is currently commercially available.
Workshop participants embraced unconventional compositions, often utilizing non-rectangular shapes and combining several panels.
Our interactive prototypes, in turn, showed that even simple setups can make such concepts tangible despite current hardware limitations.
Few foldable devices exist on the market today, and their form factors are only beginning to diversify.
This gives visualization research the unique opportunity to play a constructive role in shaping this device class.
We hope that the conceptual foundations and design concepts presented in this paper can inspire and inform the development of future visualization applications for this evolving class of devices.

\acknowledgments{
We thank P. Isenberg for commenting on early ideas and concepts, M. Hu, M. Pattikawa, and O. Taher for contributing as student assistants, and the anonymous reviewers for their valuable input.
We thank Schloss Dagstuhl - Leibniz Center for Informatics and the participants of the seminar 25082 for the opportunity to discuss early ideas of this work.

Part of this work was funded by the Deutsche Forschungsgemeinschaft (DFG, German Research Foundation) -- 545609014; by the DFG as part of Germany’s Excellence Strategy -- EXC 2050/2 -- Project ID 390696704 -- Cluster of Excellence ``Centre for Tactile Internet with Human-in-the-Loop'' (CeTI) of TUD Dresden University of Technology; as well as the German Federal Ministry of Research, Technology and Space (BMFTR, SCADS22B) and the Saxon State Ministry for Science, Culture and Tourism (SMWK) by funding the competence center for Big Data and AI ``ScaDS.AI Dresden/Leipzig''.
Part of this work was supported by the National Natural Science Foundation of China under project No. 62202397.}

\section*{Images/figures license/copyright}
\label{sec:figure_credits}
All figures in this article are and remain under our own personal copyright, with the permission to be used here.
We also share them at \href{https://osf.io/enmyu}{osf.io/enmyu} under the \href{https://creativecommons.org/licenses/by/4.0/}{Creative Commons (\ccLogo\,\ccAttribution\ \mbox{CC BY 4.0})} license.

\bibliographystyle{abbrv-doi-hyperref-narrow}

\bibliography{references_short}
%
\end{document}